# Evidence of Direct Electronic Band Gap in two-dimensional van der Waals Indium Selenide crystals

Hugo Henck[1], Debora Pierucci[2], Jihene Zribi[1], Federico Bisti[2], Evangelos Papalazarou[3], Jean Christophe Girard[1], Julien Chaste[1], François Bertran[4], Patrick Le Fevre[4], Fausto Sirotti[5], Luca Perfetti[6], Christine Giorgetti[6], Abhay Shukla[7], Julien E. Rault[4] and Abdelkarim Ouerghi[1]

[1]Centre de Nanosciences et de Nanotechnologies, CNRS, Univ. Paris-Sud, Université Paris-Saclay, C2N – Marcoussis, 91460 Marcoussis, France
[2]CELLS - ALBA Synchrotron Radiation Facility, Carrer de la Llum 2-26, 08290 Cerdanyola del Valles, Barcelona, Spain
[3]Laboratoire de Physique des Solides, CNRS, Université Paris-Saclay, Université Paris-Sud, 91405 Orsay, France
[4] Synchrotron-SOLEIL, Saint-Aubin, BP48, F91192 Gif sur Yvette Cedex, France
[5]Laboratoire de Physique de la Matière Condensée, Ecole Polytechnique, CNRS, Université Paris Saclay, 91128 Palaiseau Cedex, France
[6]Laboratoire des Solides Irradiés, Ecole Polytechnique, CNRS, CEA, Université Paris-Saclay, 91128 Palaiseau Cedex, France
[7]Institut de Minéralogie, de Physique des Matériaux et de Cosmochimie (IMPMC), Sorbonne Universités—UPMC Univ. Paris 06, UMR CNRS 7590, Muséum National d'Histoire Naturelle, IRD UMR 206, 4 Place Jussieu, 75005 Paris, France

Metal mono-chalcogenide compounds offer a large variety of electronic properties depending on chemical composition, number of layers and stacking-order. Among them, the InSe has attracted much attention due to the promise of outstanding electronic properties, attractive quantum physics, and high photo-response. Precise experimental determination of the electronic structure of InSe is sorely needed for better understanding of potential properties and device applications. Here, combining scanning tunneling spectroscopy (STS) and two-photon photoemission spectroscopy (2PPE), we demonstrate that InSe exhibit a direct band gap of about 1.25 eV located at the Γ point of the Brillouin zone (BZ). STS measurements underline the presence of a finite and almost constant density of states (DOS) near the conduction band minimum (CBM) and a very sharp one near the maximum of the valence band (VMB). This particular DOS is generated by a poorly dispersive nature of the top valence band, as shown by angle resolved photoemission spectroscopy (ARPES) investigation. In fact, a hole effective mass of about $m^*/m_0 = -0.95$ ($\overline{\Gamma K}$ direction) was measured. Moreover, using ARPES measurements a spin-orbit splitting of the deeper-lying bands of about 0.35 eV was evidenced. These findings allow a deeper understanding of the InSe electronic properties underlying the potential of III−VI semiconductors for electronic and photonic technologies.

**Keywords:** Indium Selenide – ARPES – 2PPE – STM/STS – Electronic band structure – Direct Band Gap - van der Waals materials

Metal (M) mono-chalcogenide (X) III-VI compound (MX), as InSe, GaSe, GaS and GaTe, has emerged as a promising 2D semiconductor[1–6]. Its crystal structure consists in covalently-bound "X-M-M-X" sheets, stacked vertically through van der Waals (vdW) interaction[7–9] The electronic band structure of many two-dimensional (2D) vdW materials changes as the thickness is reduced down to a few layers[10]. One well known example is the indirect to direct gap transition that occurs at monolayer thickness of Mo and W transition metal dichalcogenides (TMDs)[11]. To date, a wide variety of 2D vdW crystals such as TMDs, and hexagonal boron nitride (hBN) have been investigated, and exploited as single crystals or in combination with graphene to create functional devices[12–15].

At odds with the dichalcogenide ($MX_2$) compounds, MX (*e.g.*, GaSe and InSe) are direct bandgap semi-conductors in the bulk (2.1eV and 1.2eV) but they evolve to be indirect bandgap when thinned down to few layers[16–18]. Being optically active in the infrared or visible range is a significant advantage for 2D materials since graphene (gapless) and hexagonal boron nitride (>5eV) are not very practical for optoelectronic applications. The indirect band gap transition has not deterred the successful use of few-MLs GaSe and InSe as fast photodetectors[19–22]. This kind of indirect bandgap transition is also a subject of unusual physics, where a near flat valence band pinches down at the Γ point ('Mexican hat') and creates a Van Hove singularity close to the valence band maximum[23]. This peculiar density of states singularity could lead to interesting effects such as tunable magnetism depending on the doping of the single layer or other instabilities[24,25]. Some recent works highlight the potential applications of InSe and related III−VI 2D materials in optoelectronics[4, 18]. Tamalampudi *et al.* [19] showed that devices based on few-layered InSe obtained by mechanical exfoliation can be used as photosensor with high photoresponsivity. Additionally, electroluminescence was observed in vertically stacked InSe/GaSe heterojunction based on mechanical exfoliation method of 2D vdW materials[26]. Moreover the use of such heterojunction of layered semiconductors where the top and the bottom of the valence band are located closed to the Γ point can be very useful in the building of finely-tuned artificial semiconductors[27]. In terms of electrical transport device, it was recently showed that in a InSe field effect transistors (FET) capped with h-BN operating at room temperature, the carrier mobility can be improved up to ∼$10^3$ cm$^2$/(V s) getting close to the high Hall mobility value of the InSe bulk, which is also well above that one of the TMDs[1]. This high quality devices have also brought to the realization of gate-controlled quantum dots in the Coulomb blockade regime with few-layers InSe[28].

Although optical and transport studies of MX have made rapid progress, the intrinsic electronic structure is still not precisely understood with many open questions: do MX semiconductors have a direct bandgap at the Γ points? Fortunately, ARPES and STM/STS have the potential to answer these questions. We report here a detailed experimental study of InSe electronic band structure. ARPES was used to investigate the InSe band structure and the position of the valence band maximum relative to the Fermi level. This information coupled with the 2PPE and STS measurements gives the band offsets and the electronic band gap in our crystal. Complementary micro-Raman spectroscopy analysis was conducted to study the structural properties of the crystal exploring the vibration frequencies of phonons corresponding to the characteristic vibrational modes of InSe.

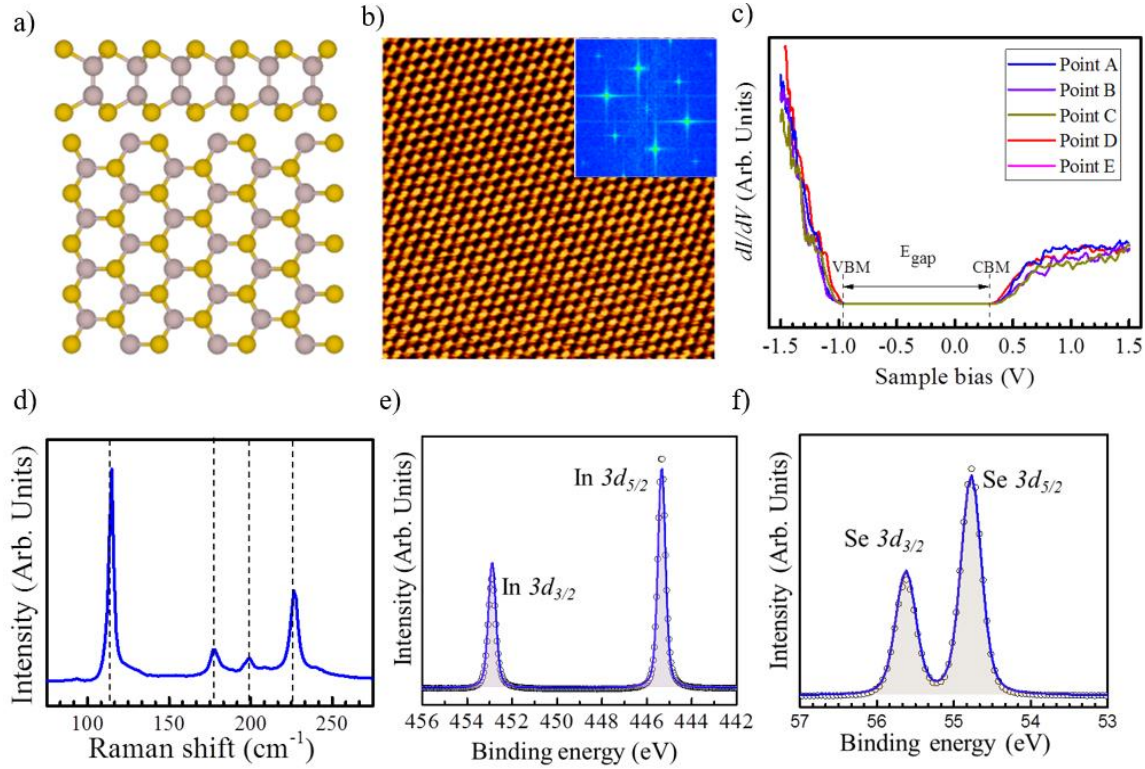

**Figure 1:** (a) A schematic diagram of the side view and top view of hexagonal structure of InSe (the grey spheres refer to indium atoms and the yellow spheres refer to the selenium atoms), b) Typical STM image (10 nm x 10 nm) of InSe ($V_{Tip}$ = 1.3 eV, $I_t$ = 400 pA, T = 4 K), c) STS d$I$/d$V$ spectrum showing the electronic bandgap of InSe, d) The characteristic micro-Raman spectrum of ε-InSe crystal. High resolution XPS of InSe at hν = 600 eV: e) In 3d and f) Se 3d.

Similarly to GaSe, InSe is a layered crystal with strong covalent in-plane interatomic bonding and weaker vdW inter-plane bonding. A single layer (ML) of InSe consists of four hexagonally arranged atoms linked via covalent bonds in the sequence Se–In–In–Se, with thickness of about 0.8 nm[5] (Figure 1 (a)). In bulk, these individual InSe layers are held together by vdW forces[29]. Here, high-quality InSe single crystals were synthesized by the flux method[30–32]. In order to study the structural and electronic properties of our sample, we carried out STM/STS measurements at low temperature of 4.2 K. An atomic lattice image, as shown in Figure 1(b), can be observed on the InSe sample. The Fourier transform (FT) of the STM images is shown in inset of Figure 1(b): the surface exhibits clearly one ordered structures with the (1x1) symmetry of the InSe. From the FT, it is possible to measure the spatial periodicity of a = 0.40 nm in the InSe sample. The resolved differential conductance (*dI/dV*) spectroscopy versus bias voltage spectra, proportional to the local density of states (LDOS), measured in different positions of the InSe sample are shown in Figure 1(c). The valence band maximum (VBM) is located at 0.96 ± 0.05 eV below the Fermi level (*i.e.*, zero bias on the *dI/dV* spectra), and the conduction band minimum (CBM) is located at 0.29 ± 0.03 eV above the Fermi Level, thereby yielding an intrinsic electron quasiparticle bandgap of $E_g$ = 1.25 ± 0.08 eV. The uncertainty in $E_g$ is the results of the lateral position variations. The relative position of $E_F$ with respect to the band edges reveals *n*-type doping for our samples which can be attributed to intrinsic point defects such as vacancies and/or lattice antisites, responsible for n-doping in other 2D materials[33–35]. The LDOS of the InSe is very sharp at the top of the valence band; near the conduction band minimum (within the range of 1.0 – 1.3 eV) it is constant. This peculiar density of state, as shown by theoretical calculations[36], should give excellent thermoelectric performance to the InSe crystal.

In order to probe the chemical and structural properties of synthetic InSe single crystal, micro-Raman and high-resolution x-ray photoemission spectroscopy (HR-XPS) investigations were performed on the same sample. The InSe micro-Raman spectrum measured at room temperature is reported in Figure 1(d). We can distinguish from this spectrum the typical vibrational modes reported previously for bulk InSe[37,38] in non-resonant conditions (laser excitation 532 nm): the peaks at 114.3, 177.8 and 226.9 cm$^{-1}$ corresponding respectively to the vibrational modes $A_{1g}^1$, $E_{2g}^1$ and $A_{1g}^2$ [4,6,20,37]. A small peak at around 200 cm$^{-1}$ is also present, signature of the ε-polytype[39–41]. HR-XPS measurements were carried out on the InSe crystals at the Cassiopée beamline of the synchrotron Soleil (France) at 50K. The overview spectrum on a wide energy range (hv = 600 eV) (Figure S1) shows the only presence of the In (In 3d and In 4d) and Se (Se 3s, Se 3p and Se 3d) peaks, without any signal related to oxygen or carbon, underlying the absence of contaminations. High resolution spectra for In (In 3d and In 4d), and Se (Se 3d) are also recorded at 600 eV (Figure 1 (e), Figure S2 and 1(f), respectively). The different components of the spectra were decomposed by a curve fitting procedure (see Supplementary Information Sec. I). The experimental data are displayed in dots and the blue solid lines represent the envelope of the fitted components. The In 3d spectrum presents two peaks at a binding energy (BE) of 445.3 eV and 452.8 eV, which can be attributed to the In 3d $_{5/2}$ and In 3d $_{3/2}$ (spin-orbit (SO) splitting of 7.5 eV, with the expected $3d_{3/2}:3d_{5/2}$ ratio of 0.66). Similarly, two components are also present for the Se 3d peak, corresponding to the Se 3d $_{5/2}$ and 3d $_{3/2}$ at 54.7 eV and 55.5 eV, respectively, with a SO splitting of 0.8 eV and the expected $3d_{3/2}:3d_{5/2}$ ratio of 0.66. No other components are detected in the spectra related to oxidized InSe (i.e., Se-O around 59 eV)[6,42–44]. The obtained binding energy values are in agreement with previously reported ones obtained for a single crystal InSe (*n*-type doping)[42,45]. Moreover, a quantitative analysis, obtained using the intensity of the In 4d and Se 3d peaks (scaled by the specific photoemission cross section, see Sec. IV of the Supplementary Information)[46–49], shows a In:Se ratio of about 1.01, confirming the monochalcogenide phase of the crystal[50].

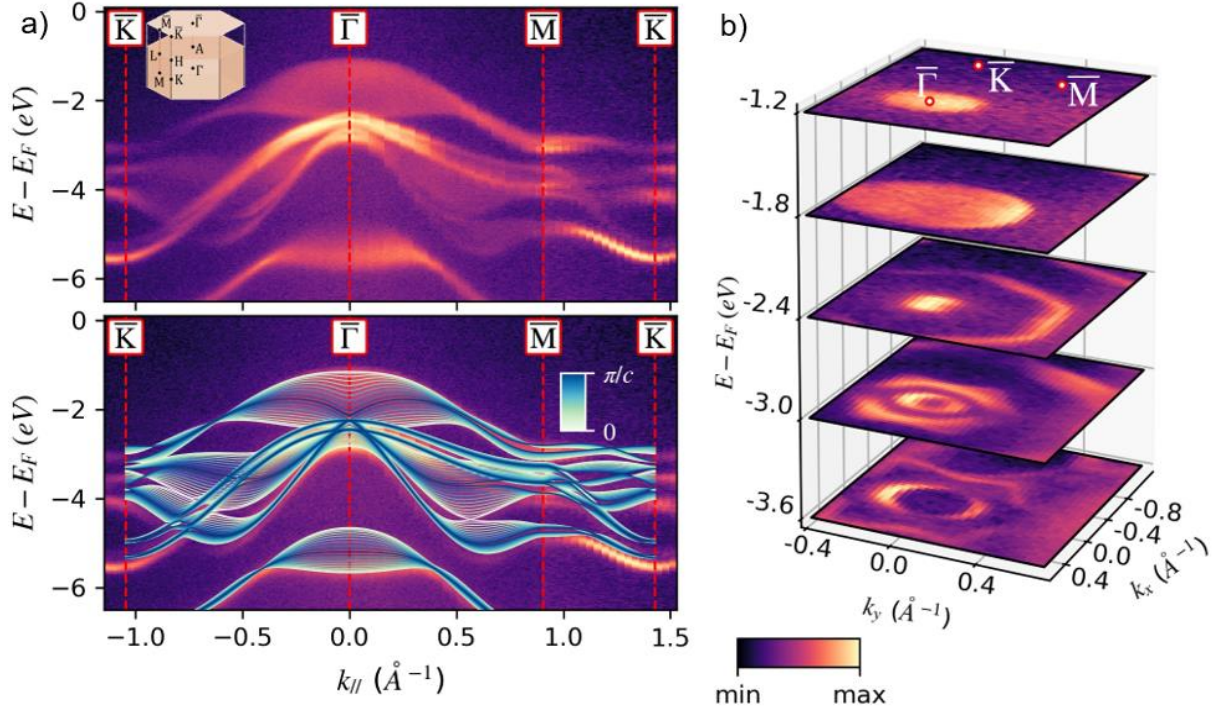

**Figure 2: a)** Top panel: High resolution map of the electronic band structure of InSe crystals collected at hv = 60 eV and T = 50 K along the $\overline{K\Gamma MK}$ direction, the Fermi Level is located at the zero of the binding energy; low panel: comparison between the experimental and the theoretical band structure of InSe in white-blue color scale, the evolution of the band structure from $k_z=0$ to $k_z=\pi/c$ is shown. **b)** Isoenergy contours along the $\overline{\Gamma KM}$ plane.

More insight into the InSe electronic structure was obtained using angle resolved photoemission spectroscopy (ARPES) combined with density functional theory (DFT) calculations (more details in supplementary information Section V). In figure 2(a) top panel, we report the InSe band structure projection on the surface Brillouin zone (BZ) (see insert figure 2(a) top panel) along the $\overline{K\Gamma MK}$, as explored by using a photon energy of 60 eV. The topmost part of the valence band is characterized by down-dispersing paraboloid centered at the $\overline{\Gamma}$ point. The VBM is located at -1.05±0.05 eV, in agreement with the STS measurements. As shown by theoretical calculations[51], this paraboloid has a $p_z$-like symmetry (out of plane orbital), which differentiates from the $p_x$ and $p_y$-like (in-plane orbitals) located at higher distance from the Fermi level (about -2.5 eV). The spin orbit interaction induces an energy splitting of these two latter bands of $\Delta_{SO}$ = 0.35 eV (see figure S3 in supplementary information)[51,52]. The large uncertainty of the $k_z$ vector, due to the low photon energy (60 eV)[53], spreads the paraboloid contours to low binding energy (see also figure 3 (d) – (f) for the $\overline{K\Gamma}$ direction explored by using the different photon energies). This effect is more pronounced around $\overline{\Gamma}$ than $\overline{K}$ or $\overline{M}$, due to the lower band dispersion along $k_z$ of the states around the latter points than $\overline{\Gamma}$. In figure 2(a) bottom panel, the calculated band structure is superimposed on the experimental data. The excellent agreement is confirming we are probing the band structure of the bulk ε-InSe. In white-blue color scale, the evolution of the band structure from $k_z$ = 0 to $k_z$ = π/c is highlighted. From the projection of the valence band on the surface Brillouin zone ($\overline{\Gamma K}$ direction) of figure

2 (a) effective mass of the hole close to the $\bar{\Gamma}$ point is calculated. The experimental dispersion has been fitted with a parabolic model $E(\mathbf{k}) = E_0 + \frac{\hbar^2}{2m^*}\mathbf{k}^2$ where m* is the effective electron mass and ℏ is the reduced Planck constant. We found that the hole effective mass at the $\bar{\Gamma}$ point is about m*/m$_0$ = -0.95 ± 0.05. One can notice that this mass is close to m$_0$ which implies that the InSe band is poorly dispersive. This feature leads to the appearance of a very sharp DOS near the top of the valence band of InSe as observed by STS.

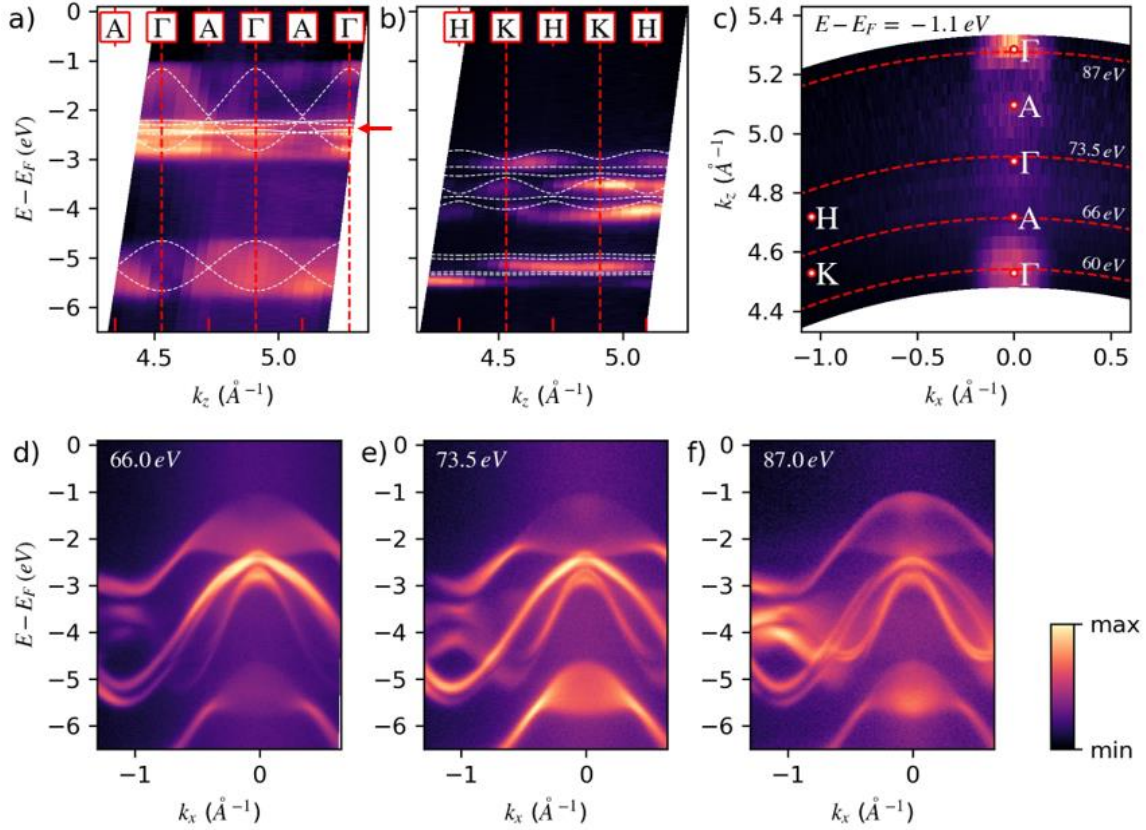

**Figure 3:** Photon energy scan (from 58.5 to 88.5 eV): a) band dispersion along ΓA, b) band dispersion along KH and c) isoenergy contour along the ΓAK plane. The theoretical calculations are represented by white dashed lines. Electronic band structure of InSe along the ($\overline{\Gamma K}$) high-symmetry direction collected at different photon energy (hv = 66, 73.5 and 87 eV) and T = 50 K. Following the figure 3 (c), the reported photon energies are the ones passing respectively through the A point, and two Γ points.

The isoenergy contours in the $\overline{\Gamma KM}$ plane (figure 2 (b)) confirm the presence of states only around the $\bar{\Gamma}$ point. They show an in-plane isotropic extension of the paraboloid at least until -1.8 eV, reaching a hexagonal shape once approaching the corners of Brillouin zone at lower binding energy. The in-plane orbitals, instead, establish a star-shaped contour following the surface symmetry. In figure 3 (a) and (b) the out of plane direction is explored by a photon energy scan (from 58.5 to 88.5 eV), giving access to the $k_z$ vector. The band dispersion along ΓA (figure 3(a)) shows a broad parabola around Γ points for the top most part of the valence band (mostly visible around $k_z$ = 4.5 Å$^{-1}$) and an almost non-dispersing features around -2.5 eV (red arrow). These observations are in excellent agreement with the theoretical calculations (white dashed lines in the figure 3), confirming also the previous assignment of an out-of-plane $p_z$-like orbital for the topmost part of the valence band, and an in-plane $p_x$ and $p_y$- like orbitals at around -2.5 eV. An additional intensity modulation, following the perpendicular periodicity of the crystal, is also recognizable in the broad band centered at -5.0 eV, suggesting an out-of-plane orbital character for this band. The band dispersion along KH (figure 3(b)) is instead showing almost non-dispersing features as expected from calculation (white dashed lines), reflecting the in-plane 2-dimentional character of the bands around K and H points. To complete the investigation we also show the isoenergy contour in the ΓAK plane (Figure 3 (c)) near the valence band maximum. This contour confirms the location of the VBM around the Γ points, in fact almost no signal is present in other places of the BZ. For the photon energies of hν = 66 eV, 73.5eV and 87 eV we are passing through the A point, and two Γ points, respectively. The corresponding band structures for these photon energies are reported in figure 3 (d), (e) and (f). The topmost part of the valence band near the A point, shows a reduced signal at the center of $k_x$ in favors of an increase on the sides. The signal almost vanishes for the Γ point at 73.5 eV. This reduced signal intensity for the Γ point in the middle of the explored $k_z$ range (around $k_z$ = 4.8 Å$^{-1}$), as recognizable also in figure 3 (a) and (c), is due to an interference ARPES effect induced by the multilayer structure of the crystal[54–56]. Therefore it can be considered as a signature of 3-dimentionality for the band structure, but also as an indication of good quality for the investigated crystal. Finally, the Γ point explored with the highest photon energy (87 eV) shows a reduced spread of the paraboloid contours to low binding energy, due to the reduced uncertainty on the $k_z$ vector[53].

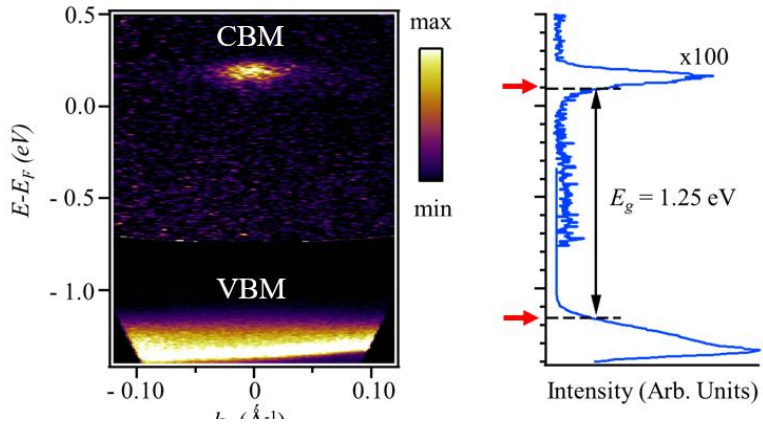

**Figure 4:** 2PPE image acquired along $\overline{\Gamma K}$ direction using a 6.24 eV of probe and 1.56 eV of pump photon energy. An electronic band gap of 1.25 eV at $\overline{\Gamma}$ point was measured. Red arrows show VBM and CBM obtained from an energy dispersion curve around $\overline{\Gamma}$.

A complete picture of the electronic structure of the InSe crystal is obtained using 2PPE spectroscopy giving access the unoccupied state. We used a Ti:sapphire laser system delivering 6μJ pulses with a repetition rate of 250 kHz. The system was photo-excited using the fundamental beam, ω = 1.56 eV (pump), and probed using the 4ω pulse of 6.24 eV. Figure 4(a) and (b) shows a 2PPE image acquired along $\overline{\Gamma K}$ direction. In the image we can observe that the conduction band minimum (CBM) in InSe crystals is located at the Γ point, as the VBM. A single-particle electronic bandgap of $E_g = E_{CBM} - E_{VBM} = 1.25 \pm 0.04$ eV was measured (figure 4 (b)), in agreement with the STS spectra. This result suggests that InSe has a direct electronic bandgap at Γ point. In accord, pump-probe dynamics performed in this system have shown a dispersion of unoccupied states with CBM in the center of Brillouin zone[31].

In summary, we have presented an exhaustive study of the electronic band structure of the InSe crystal combining spectroscopic tools and theoretical investigation (DFT). The joint analysis of the DOS using STS and of the band structure using 2PPE indicates that the InSe crystal presents a direct quasiparticle band gap of about 1.25 eV located at the Γ point of the BZ. ARPES spectra reveal that the spin-orbit interaction lifts band degeneracies in deeper lying valence band, causing a splitting of about 0.35 eV. These properties holds greats potential for a wide range of applications of vdW heterostructures based on layered InSe, ranging from thermoelectric applications to nanoelectronic and optoelectronic ones.

**Acknowledgements:** We acknowledge funding from the French National Research Agency ANR H2DH (ANR-15-CE24-0016), RhomboG (ANR-17-CE24-0030), Labex NanoSaclay (ANR-10-LABX-0035) and EU-H2020 research and innovation program under grant agreement no. 654360, NFFA-Europe. Computer time has been grantedby GENCI (Project No. 544).

**Competing financial interests:** The authors declare no competing financial interests.

# Supporting Information

# Evidence of Direct Electronic Band Gap in two-dimensional van der Waals Indium Selenide crystals

Hugo Henck[1], Debora Pierucci[2], Jihene Zribi[1], Federico Bisti[2], Evangelos Papalazarou[3], Jean Christophe Girard[1], Julien Chaste[1], François Bertran[4], Patrick Le Fevre[4], Fausto Sirotti[5], Luca Perfetti[6], Christine Giorgetti[6], Abhay Shukla[7], Julien E. Rault[4] and Abdelkarim Ouerghi[1]

[1]Centre de Nanosciences et de Nanotechnologies, CNRS, Univ. Paris-Sud, Université Paris-Saclay, C2N – Marcoussis, 91460 Marcoussis, France
[2]CELLS - ALBA Synchrotron Radiation Facility, Carrer de la Llum 2-26, 08290 Cerdanyola del Valles, Barcelona, Spain
[3]Laboratoire de Physique des Solides, CNRS, Université Paris-Saclay, Université Paris-Sud, 91405 Orsay, France
[4] Synchrotron-SOLEIL, Saint-Aubin, BP48, F91192 Gif sur Yvette Cedex, France
[5]Laboratoire de Physique de la Matière Condensée, Ecole Polytechnique, CNRS, Université Paris Saclay, 91128 Palaiseau Cedex, France
[6]Laboratoire des Solides Irradiés, Ecole Polytechnique, CNRS, CEA, Université Paris-Saclay, 91128 Palaiseau Cedex, France
[7]Institut de Minéralogie, de Physique des Matériaux et de Cosmochimie (IMPMC), Sorbonne Universités—UPMC Univ. Paris 06, UMR CNRS 7590, Muséum National d'Histoire Naturelle, IRD UMR 206, 4 Place Jussieu, 75005 Paris, France

## I. Methods and characterization:

Scanninc tunnenling microscopy and spectroscopy (STM/STS) measurements:

STM/STS measurements were carried out using an Omicron ultra-high vacuum low temperature scanning tunnelling microscope (UHV-LT-STM). The STM tips used in this study were electrochemically-etched polycrystalline tungsten tips, flashed by Joule heating under UHV to temperatures up to white light emission followed by field emission. This procedure allowed to estimate the tip apex radius but also to control the stability of the emission current relative to the quality and cleanness of the tip. In addition during STM/STS experiments, the spectroscopic behavior of the tips were systematically checked on a clean Au(111) surface in order to ensure that the tips were free of adventitious carbon contaminations. STM/STS were acquired at 4.2 K in the constant current mode for different bias voltages V applied to the sample. For the STS measurements, performed at $T = 77$ K, the I(V) characteristics were acquired while the feedback loop was inactive, the differential conductivity $dI/dV$ (V, x, y), proportional to the LDOS, was measured directly by using a lock-in technique. For this purpose a small AC modulation voltage $V_{mod}$ was added to $V$ ($V_{mod,p-p}$=10 mV, $f_{mod}$= 973 Hz) and the signal $dI$ detected by the lock-in amplifier was used to determine the differential conductivity $dI/dV_{mod}$. The surface of the crystal was prepared by *in situ* cleaving in UHV. The electronic band gap value for the InSe crystal, $E_g = 1.25 \pm 0.08$ eV, was obtained through a statistical analysis of several STS curves obtained on several positions of the sample.

Raman measurements:

The Raman spectra measurements were performed using a confocal commercial Renishaw micro-Raman microscope with a 100× objective and a Si detector (detection range up to ~2.2 eV) using a 532 nm laser in an ambient environment at room temperature. To ensure the reproducibility of the data, we followed a careful alignment and optimization protocol. In addition, the excitation laser was focused onto the samples with spot diameter of ~1 μm and incident power of ~5 mW. The integration time was optimized to obtain a satisfactory signal-to-noise ratio.

Photoemission spectroscopy:

X-ray Photoemission Spectroscopy (XPS)/Angle Resolved Photoemission Spectroscopy (ARPES) experiments were performed in ultra-high vacuum at the 3rd generation SOLEIL Synchrotron facility (Saint-Aubin, France) on the CASSIOPEE beamline. The CASSIOPEE beamline is equipped with a Scienta R4000 hemisperical electron analyzer whose angular acceptance is ±15° (Scienta Wide Angle Lens). All the experiment were done a T= 50 K. The total angle and energy resolutions were 0.25° and 16 meV (100 meV for the XPS spectra). The mean diameter of the incident photon beam was smaller than 50 *μ*m. A for the STM/STS measurement the crystal was prepared by *in situ* cleaving in UHV. For the fitting procedure of the XPS data a Shirley background was subtracted in all core level spectra. The Se 3d and In 4d and 3d spectra were fitted using sums of Voigt curves composed by the convolution of a Gaussian and a Lorentzian line shapes with a FWHM ~ 0.4 eV.

## II.   X-ray photoemission spectroscopy (XPS) measurements:

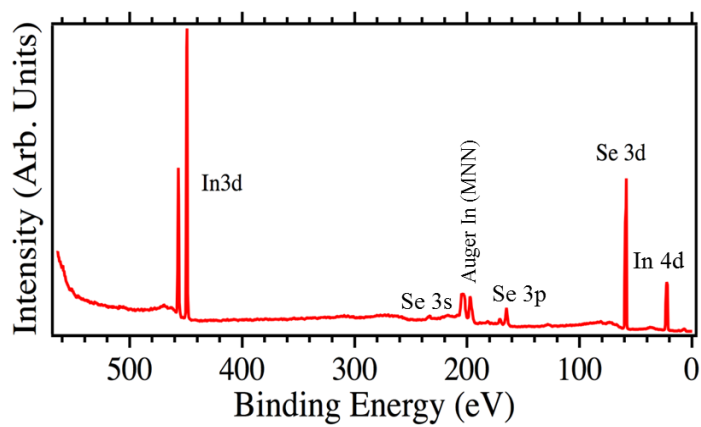

**Figure S1:** High resolution wide XPS spectrum of InSe measured at hv = 600 eV.

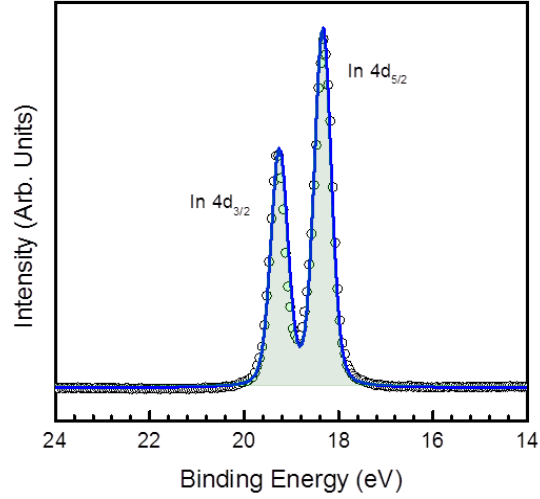

**Figure S2:** High resolution XPS spectrum of In 4d measured at hν = 600 eV.

As for the Se 3d and In 3d the high resolution spectrum of In 4d was acquired using a photon energy of 600 eV. The experimental data points are displayed in dots and the blue solid lines represent the envelope of the fitted components. The In 4d spectrum presents two peaks at a binding energy (BE) of 18.3 eV and 19.4 eV, which can be attributed to the In 4d $_{5/2}$ and In 4d $_{3/2}$ (spin-orbit (SO) splitting of 0.9 eV and  3d $_{3/2}$:3d $_{5/2}$ ratio of 0.66 ) [1,2].

### III.   Quantitative surface analysis by XPS:

In order to calculate the atomic concentration of In and Se in our sample, the following equation (1) is used[3–5]:

$$C_A = \frac{I_A}{F_A} \qquad (1)$$

with $I_A$ is the photoemission intensity for a specific orbital from element A, in this case the In 4d and Se 3d and $F_A$ is the relative sensitive factor of the corresponding element A = In or Se:

$$F_A = \sigma_A(h\nu) \cdot T\,(Ek_A) \cdot \lambda(Ek_A) \qquad (2)$$

where:

- $\sigma_A(h\nu)$ is the element, orbital and photon energy ($h\nu$) specific photoemission cross section[6]
- $T\,(Ek_A)$ is the transmission function of the spectrometer at kinetic energy $Ek_A$, that can be described as :
  $$T\,(Ek_A) \sim Ek_A^{\alpha} \quad (\alpha \text{ depends on the particular spectrometer used})$$

- $\lambda(Ek_A)$ is the kinetic energy dependent electron mean free path that can be approximated as
  $$\lambda(Ek_A) \sim Ek_A^{\beta} \quad \text{with}^{7} \; \beta \sim 0.5$$

Then if we compare the intensity $I_{In\ 4d}$ and $I_{Se\ 3d}$ the ratio of In and Se is therefore:

$$\frac{C_{In}}{C_{Se}} = \frac{I_{In\ 4d}}{I_{Se\ 3d}} \cdot \frac{\sigma_{Se\ 3d}(h\nu)}{\sigma_{In\ 4d}(h\nu)} \cdot \left(\frac{Ek_{Se\ 3d}}{Ek_{In\ 4d}}\right)^{\alpha+\beta} \qquad (3)$$

In 4d and Se 3d have close binding energy values. Then using the same photon energy $h\nu = 600$ eV, they present also close kinetic energy values that means $\left(\frac{Ek_{Se\,3d}}{Ek_{In\,4d}}\right)^{\alpha+\beta} \sim 1$. Then:

$$\frac{C_{In}}{C_{Se}} \sim \frac{I_{In\,4d}}{I_{Se\,3d}} \cdot \frac{\sigma_{Se\,3d}(h\nu)}{\sigma_{In\,4d}(h\nu)} \qquad (4)$$

### IV. Angle resolved photoemission spectroscopy (ARPES) measurements:

The spin-orbit coupling, although does not break spin degeneracy at the $\bar{\Gamma}$ point, does lift band degeneracies in deeper lying valence band, causing a splitting of about 0.35 eV, which is visible in the ARPES spectrum (Figure S3)

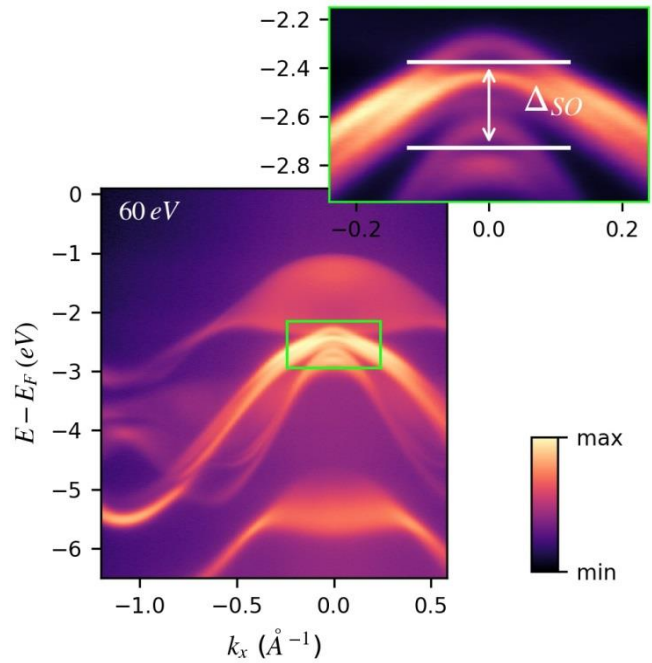

**Figure S3:** High resolution map of the electronic band structure of InSe crystal along the $\overline{\Gamma K}$ high-symmetry direction collected at h$\nu$ = 60 eV and T = 50 K. The band located around -2.5 eV due to spin orbit interaction are separated in energy of about $\Delta_{SO}$ = 0.35 eV (zoom in insert)

### V. Ab initio calculation of InSe band structure

The three-dimentional ε-polytype of InSe (space group $D_{3h}^1$ in Schoenflies notation) has been build using the lattice parameters a=3.9553 Å, d$_{M-M}$= 2.741 Å, d$_{X-X}$= 5.298 Å and c=16.64 Å and AB stacking [8–13]. These values are very closed to experimental ones and no influence of such a small difference is expected.

The band structure has been calculated within Density Functional Theory in the Local Density Approximation (DFT-LDA) using the ABINIT code[14]. We used for In and Se norm-conserving Troullier-Martins pseudopotentials generated by fhi98PP. ε -InSe contains 8 atoms per unit cell. Each

In atom participates with 13 valence electrons $[4d^{10}5s^{2}5p^{1}]$, and each Se atom with 6 $[4s^{2}4p^{4}]$. The electronic density has been converged on a grid of 8 x 8 x 2 **k**-points, with an energy cutoff of 90 Ry. The band structure was calculated with 151 **k**-points along KΓKM, and the corresponding path has been repeated for 10 cuts along the k$_z$ direction to obtain the surface projected band structure. 9 **k**-points were used to follow ΓA and HK dispersions.